# The Road Less Traveled: Non-traditional Ways of Communicating Astronomy with the Public[1]

Michael J. West

Dept. of Physics & Astronomy, University of Hawaii, 200 W. Kawili Street, Hilo, Hawaii, 96720, USA; e-mail: westm@hawaii.edu

**Abstract**

In an age of media saturation, how can astronomers succeed in grabbing the public's attention to increase awareness and understanding of astronomy? Here I discuss some creative alternatives to press releases, public lectures, television programs, books, magazine articles, and other traditional ways of bringing astronomy to a wide audience. By thinking outside the box and employing novel tools – from truly terrible sci-fi movies, to modern Stonehenges, to music from the stars – astronomers are finding effective new ways of communicating the wonders of the universe to people of all ages.

**Introduction**

As a number of talks at this conference have emphasized, those of us involved in astronomy education and outreach face an uphill struggle to reach the public through books, press releases, images, and other traditional ways of communicating astronomy. Let me illustrate the challenges with two specific examples:

1) In July 1969, Neil Armstrong became the first person to walk on the Moon, an event that surely ranks as one of the greatest achievements in human history. Just three and a half years later, in January 1973, Elvis Presley's "Aloha from Hawaii" television special was beamed via satellite from Honolulu to over 40 countries around the world. Guess which of these two events was watched by more television viewers? It is estimated that more than a billion people around the world tuned in to watch the "Hunk o' Burning Love" on television, compared to about 600 million people who viewed Armstrong's walk on the Moon. When the Elvis special was shown on American television, more Americans watched it than watched the Apollo 11 astronauts walk on the Moon. It's a sobering reminder to those who wish to use television as a medium for communicating astronomy to the public that they face daunting competition from soap operas, reality television shows, sports, sitcoms, and other popular programming that captures the vast majority of television viewers.

2) Here's a depressing fact: according a recent survey by the U.S. National Endowment for the Arts, 43% of adult Americans read no books last year. Although that might explain a lot about the last U.S. presidential election, if you're a science writer it's hard not to lose the will to live, because it means that you don't have a prayer of reaching those people, no matter how clever or informative your writing might be. It's not that those 43% didn't read any

---

2science book last year -- they didn't read any book last year.  And the reading rate is declining most rapidly for young people between the ages of 18 to 24.

The late, great Richard Feynmann summarized the situation eloquently when he lamented, "Is no one inspired by our present picture of the universe?  The value of science remains unsung by singers; you are reduced to hearing not a song or poem, but an evening lecture about it.  This is not yet a scientific age."

Given this reality, how can we effectively communicate astronomy to a public that often seems uninterested in science, frequently finds it difficult, and is becoming increasingly difficult to reach via traditional media such as print and television?

My goal here is to share some creative, non-traditional ways that are being used with great success to communicate astronomy to a wide audience.  I don't pretend to be an expert on this topic, and the information that I will present here is by no means a complete survey.  Rather, it is just a partial list that will hopefully inspire others to find their own new ways of bringing astronomy to the public.

**Communicating astronomy through culture**

A key issue in astronomy outreach has always been how to address the general public's implicit question of "Why should I care about this?"  To motivate people to want to learn more about astronomy, we need to make the subject interesting, understandable, and relevant to their lives.

Culture offers a powerful "hook" to get people interested in astronomy and to make it more relevant to them.  Every one of us comes from a place with a culture, with creation stories, and with legends about the heavens.  Astronomers can use these cultural connections as a starting point to communicate the wonders of astronomy to audiences everywhere.

One place where science and culture come together is the Maunakea Discovery Center, a $27 million dollar center scheduled to open in Hilo, Hawaii in late 2005.  Conceived as a unique interpretive facility to increase public awareness of the universe and our place in it, the Maunakea Discovery Center is a community project in the truest sense.  With involvement from the astronomical community in Hawaii, the University of Hawaii, and the native Hawaiian community, the center's content will reflect the reverence for Mauna Kea shared by Hawaiians and astronomers.

The goals of the Maunakea Discovery Center are multifold, but there are two primary objectives.  First, to weave together two seemingly disparate stories – one about astronomy and one about Hawaiian culture – into a compelling story of human exploration.  Second, to provide a personal connection to science and culture.  Why should visitors to the center care about astronomy?  Why should they care about Hawaiian culture?  By tapping into universal themes that transcend science, culture and places of origin, it is possible to create a center with a broad appeal.

Why build a world-class astronomy education center in Hilo, a small town of only about 45,000 people?  Mauna Kea, which is only about an hour drive from Hilo and dominates the local landscape, is home to one of the greatest collections of telescopes in the world.  Many people – local residents and tourists alike – have heard of Mauna Kea, and many are curious to learn more about the observatories on the "White Mountain."  An estimated 100,000 visitors a year come to Mauna Kea, although most do not go to the summit to see the telescopes because of the hardships of high altitude.  With more than a million tourists visiting the Island of Hawaii each year, the Maunakea Discovery Center has a tremendous opportunity to teach countless people about the exciting astronomical discoveries being made nightly by the dozen observatories on Mauna Kea.

In addition to the scientific draw, the mountain is also a sacred site to many native Hawaiians and throughout much of Polynesia, so there is a cultural connection to Mauna Kea as well.  Hawaiian culture is not just a thing of the past; it is a living, vibrant culture that is experiencing a renaissance today, and the center aims to increase public awareness and understanding of the rich culture of the kanaka maoli (indigenous people) of Hawaii.

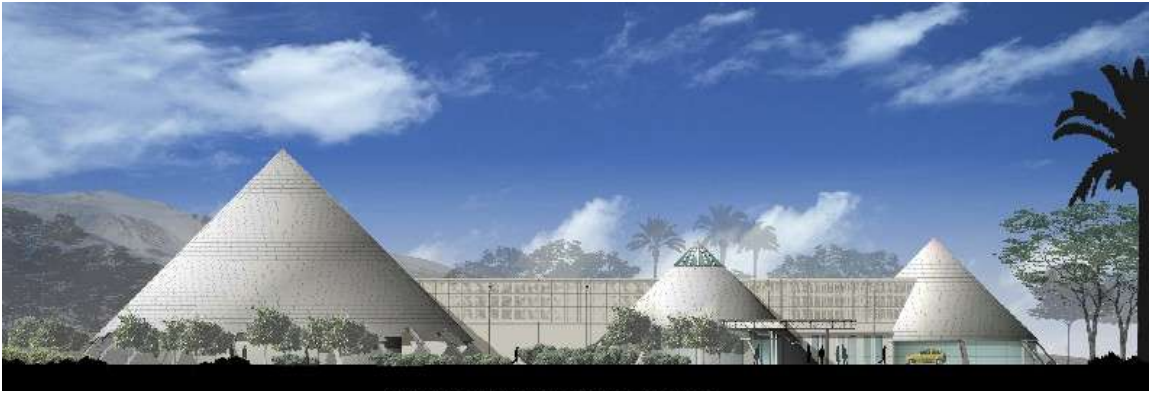

Figure 1.  Artist's rendering of the soon-to-be-completed Maunakea Discovery Center in Hilo, Hawaii.  Image courtesy of the Maunakea Discovery Center and Durant Media Five.

Visitors to the Maunakea Discovery Center will encounter interactive exhibits, panels to read, a 3-D theater, a planetarium, and live demonstrations and performances. The center's exhibits are housed in four main areas that focus on the following themes:

> **Piko.**  The center's introductory space is intended to be a place of wonder, where visitors first connect with Mauna Kea's special nature.  Piko is the Hawaiian word for umbilical cord, and in Hawaiian lore Mauna Kea is the naval of creation, a place that connects Hawaiians and their culture back through time to their ancestral origins as descendents of the gods.  This opening space attempts to recreate the feeling of enchantment that one feels on the summit of Mauna Kea, with the aim of piquing the visitor's curiosity.

> **Origins.**  The next space that visitors enter asks a basic question that people of all cultures have pondered and attempted to answer throughout history:  Where do we come from?  It begins with the Kumulipo chant, which tells an ancient Hawaiian legend of our cosmic origins, and then moves into our modern astronomical creation story, from the birth of the universe in the Big Bang to the origin of stars, planets, galaxies and life. The goal is not to compare and contrast these two very different creation stories, or to pass judgment, but simply to show them as different expressions of the ages old human quest to understand our cosmic origins.

**Voyages.** This area of the center asks: Where have we been? Where are we going from here? Here "we" can be many different groups – we astronomers, we Hawaiian people, we human beings, and the voyage may be literal or metaphorical. Visitors will learn what it was like to sail aboard a Polynesian voyaging canoe across thousands of mile of open ocean, navigating by the stars, the wind, the clouds, and other natural signposts. Visitors will also get a chance to try their hand at astronomical observations in an observatory simulator that will give a sense of what it is like to be an astronomer on a voyage of discovery into the unknown universe. Emphasis is on modern astronomers and ancient navigators are kindred spirits, driven by the same spirit of exploration. The same curiosity to find out what lies beyond the horizon that inspired the ancient Polynesians to voyage thousands of miles across the Pacific Ocean is what motivates astronomers today to explore the cosmic ocean to learn about distant worlds.

**Voices.** In the closing space of the center, visitors are given the opportunity to express their own views about Mauna Kea, culture, science, and what it all means to them. They will also have the chance to hear what previous visitors from around the world have said.

By using science and culture as partners, the Maunakea Discovery Center has an opportunity to create one of the premier astronomy interpretive centers in the world. A place that will help inspire children growing up in Hawaii to reach for the stars and to embrace the cultural heritage of the islands. A place where local residents and visitors from far away will come together to learn about our place in the cosmos. Ideally, those who come to the Maunakea Discovery Center to learn about astronomy may also leave with more awareness and appreciation of Hawaiian culture. And those who come because they are interested in Hawaiian culture may also find that their interest in astronomy is sparked. More information about the Maunakea Discovery Center can be found at www.maunakea.hawaii.edu

Of course, Hawaii isn't the only place where people have gazed into the starry skies and wondered about their origins. Another place that has successfully linked astronomy and culture is Stonehenge Aotearoa, which is located outside Wellington, New Zealand. "Aoetearoa" is the Maori name for New Zealand. Built by a very dedicated group of amateur astronomers, the goal was to create a modern Stonehenge – not an exact replica of the one in England – but one that was constructed from modern materials and which reflects the lore of the sky in the southern hemisphere and the local Maori culture.

Stonehenge Aotearoa's clever idea is to capitalize on the fact that many people – whether or not they are interested in astronomy per se – are fascinated by ancient stone circles found around the world.  It's a magnet to bring people in.  And once they are there, it's possible to teach them about the motions of the Sun and stars throughout the year, about Maori culture, and about the legendary ability of the ancient Polynesians to sail across the vast Pacific Ocean to New Zealand's shores by observing the stars.  "Stonehenge Aotearoa is a fabulous teaching tool combining art, history, different cultures around the world, star lore and astronomy," says Jennifer Picking of New Zealand's Phoenix Astronomical Society, which built the structure with a government grant and over 11,000 hours of volunteer effort.

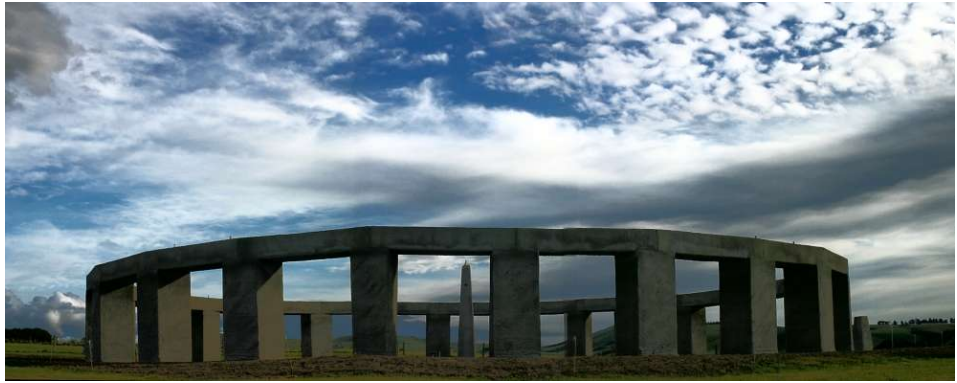

Figure 2.  Stonehenge Aotearoa, shown here, is a modern version of the ancient Stonehenge located on England's Salisbury Plain. Image courtesy of, and copyright by, Chris Picking.

Stonehenge Aotearoa opened in February 2005, and has already proven to be a popular destination for visitors.  It has a bright future as an innovative way of communicating astronomy by blending the ancient and the modern.  More information about Stonehenge Aoetearoa can be found at www.astronomynz.org.nz/stonehenge/.

**Communicating astronomy through new technology**

Another way of communicating astronomy to a wide audience is to take advantage of emerging technologies.  There are many examples of this, such as CD-ROMs, webcasts, virtual tours, and portable planetaria, all of which have made it possible to bring astronomy to the public rather than requiring the public to come to museums, observatories, or other fixed locations.  However, today's cutting-edge technology is often tomorrow's technological antique, and so those involved in astronomy outreach must be prepared to adapt quickly to opportunities presented by new technologies.

One of the newest technologies with great potential for astronomy outreach is podcasting.  Whereas just a year or two ago web logs or "blogs" were all the rage, allowing anyone to post their thoughts on the internet for others to read, the growing popularity of small, personal listening devices such as iPods has now made it easy to disseminate audio broadcasts over the internet for interested listeners.  Such "podcasts" have rapidly become the technology du jour.  Podcasting has been described as the next generation of radio, allowing listeners to find, download, and subscribe to any of thousands of free audio broadcasts that can be automatically delivered via the internet to the listener's portable audio player or personal computer.  Anyone can create a podcast with only a modest investment of resources, and if the podcast is innovative then it has the potential to reach a huge number of listeners.

With this motivation, a group of astronomy enthusiasts named Aaron Price, Pamela Gay and Travis Searle quickly capitalized on this new technology to create the first astronomy-themed podcast, which they call Slacker Astronomy.  Every week, they produce a five to ten minute podcast that discusses the latest astronomical news, along with occasional interviews, chat shows, and more.  None of the Slacker Astronomy team gets paid for this, they do it because "We believe podcasting provides a unique mechanism to extend astronomy outreach to the younger generations and to those who lead busy lifestyles."  The shows have an irreverent sense of humor that makes them both entertaining and informative.

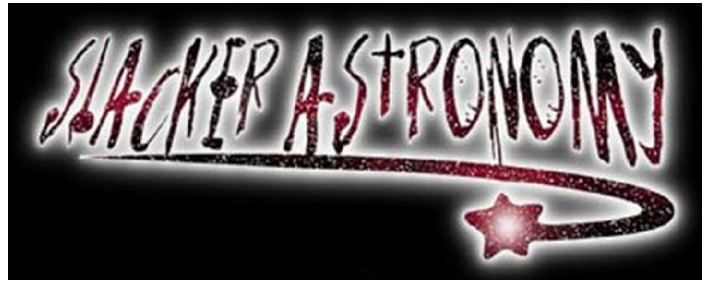

Figure 3.  Slacker Astronomy has been podcasting since February 2005 and already has nearly 10,000 listeners to its weekly shows.   Logo courtesy of Slacker Astronomy.

From its humble beginning in February 2005, Slacker Astronomy has quickly grown to over 9,000 listeners in August 2005, and that number will undoubtedly continue to rise.  These are impressive figures, and demonstrate the great potential for reaching large audiences through this new medium.  According to Aaron Price, such podcasting "is most effective as part of a greater effort in using new media.  The grand strategy is to use the blogosphere, instant messaging, online portals, podcasts, etc. to get the word out online."

The folks at Slacker Astronomy are also eager to help promote astronomical research by creating podcasts on demand.  Just send them a press release that you'd like publicized, give them two weeks notice, and they'll create a special Slacker Astronomy podcast, free of charge.  They honor embargo dates, and also allow final script approval.  It's a wonderful opportunity to reach the public in a new way, and shows what a difference a few dedicated individuals can make.  More information about Slacker Astronomy can be found on their website at www.slackerastronomy.org.

**Communicating astronomy through music**

A number of talented people are combining their passions for astronomy and music as a way of bringing astronomy to the public.  Here are a few examples:

- Jim Webb is an astronomer at Florida International University.  He does researcher in the field of active galaxies such as quasars and blazers, and he recently organized an international conference on that topic.  He's also a guitar player with a home recording studio, and to date he has recorded three astronomy-themed CDs that he gives away as gifts or sells at cost. "The songs deal with space travel, life in the universe, black holes, and our place in the cosmic scheme of things," he says. To date his CDs have traveled as far as Hungary, Sweden, Mexico and California. More information about Jim and his music can be found on his website at www.fiu.edu/~webbj/ASTROMUS.HTM

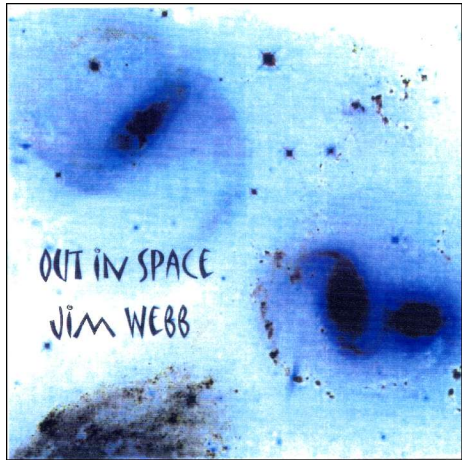

Figure 4. The cover of Out in Space, one of three CDs of astronomy-inspired music that astronomer Jim Webb has recorded at home. Image courtesy of Jim Webb.

- Fiorella Terenzi, is an astrophysicist, author and musician who has been described as "a cross between Carl Sagan and Madonna." Her CD, Music from the Galaxies, takes VLA and Westerbork radio observations of the galaxy UGC 6697, converts the frequency and intensity of the radio signal into sounds audible to the human ear, and then blends these with computer-generated music to create a sound that Time magazine has described as "part New Age, part Buck Rogers sound track, played on an oscilloscope." More information is available from her web page at www.fiorella.com

- Astrocapella is a project that Sky & Telescope magazine has described as "an astronomy class set to music." Two Astrocapella CDs have been released to date containing original, astronomically-correct a cappella songs written and performed by a group of astronomers and educators who call themselves The Chromatics. The group's philosophy is that music can serve as a powerful memory aid, and so when one hears a catchy song with astronomy-related lyrics, it stays in the listener's brain. More information can be obtained from their website at www.astrocappella.com.

- In case you're thinking that a cappella astronomy songs might not appeal to teenagers, have no fear – there's always MC Hawking. Created by an American web developer named Ken Leavitt-Lawrence, his humorous premise is that Stephen Hawking has dual careers as a famous theoretical astrophysicist and a rap star. MC Hawking's CD, A Brief History of Rhyme, consists of rap songs sung by a synthesized voice that sounds remarkably similar to the one used by the real Stephen Hawking. The MC Hawking website features a biography of MC

Hawking that blends fact and fiction, and includes digitally manipulated photos that purport to show the "Hawkman" hanging out with fellow rap artists, and a police mug shot to help establish the astrophysicist's street credibility as one tough gangsta rapper.  But it's not all just for laughs; beneath the humorous approach there really is an underlying foundation of science, with songs about such topics as entropy, unified field theory, and the Big Bang.  And in case you're thinking that this is in poor taste, don't worry.  The real Stephen Hawking has said that he is "flattered, as it's a modern day equivalent to Spitting Image."  MC Hawking creator Ken Leavitt-Lawrence says "Despite the fun I'm having with it, I have tremendous respect for the man."  More information, including a video for the MC Hawking song What We Need More of is Science, can be found at www.mchawking.com.

**Communicating astronomy through movies**

One of the earliest examples of cinematic fantasy was a pioneering short French film titled Le Voyage dans la Lune (A Trip to the Moon), based loosely on a Jules Verne story.  In this 1902 silent movie, several astronomers travel to the Moon in a spacecraft fired from a giant cannon, rough up a few Selenites that try to capture them, and then return safely to Earth.  It's a wonderful film, and was quite successful for its time.

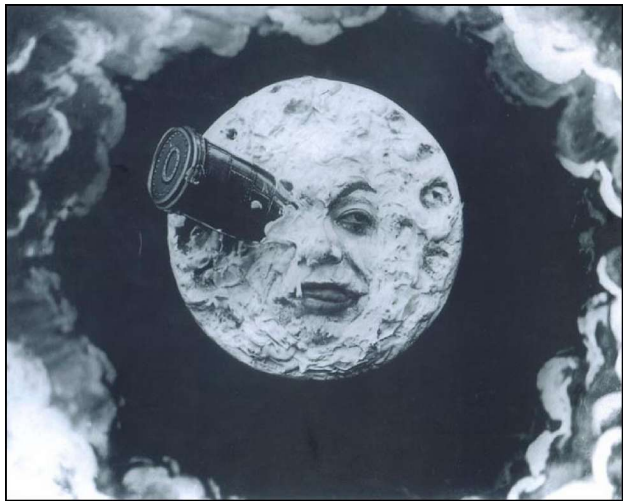

Figure 5.  A scene from the 1902 silent film A Trip to the Moon, which was written and directed by Georges Méliès.  This 14-minute film was one of the earliest science fiction movies, showing the great appeal of astronomy-related themes in cinema.

Astronomy has always been an attractive subject for movies.  Of the top five most successful movies of all time in the United States, three have themes related in one way or another to astronomy and space exploration: Star Wars, E.T The Extraterrestrial, and Star Wars: Episode I.

Capitalizing on the popularity of science fiction movies, the Harvard-Smithsonian Center for Astrophysics (CfA) in Cambridge, Massachusetts, has hit upon a novel way of using them as a tool for teaching astronomy to the public.  Sci-Fi Movie Nights is a monthly community outreach program based on the premise that "everything we know about science we learned from the movies." It is the brainchild of David Aguilar, who is Director of Public Affairs at the CfA, and it grew out of his frustration with the inaccurate way that science is often portrayed in the popular media. "Science-fiction movies have always been a motivator in sparking the interest of younger people," says Aguilar, "so why not use an attention grabbing, off-the-shelf commodity to further the teaching of science?"

The CfA shows the movies for free, and also provides free popcorn, free prizes, and free scientific discussion. Each movie is preceded by a brief introduction that explains why the movie was made, who is in it, great scenes to look for, how the special effects were done, and so on. "After the movie, the discussion now examines the science," says Aguilar. "Good science, science and technology to come, misconceptions and the bad science. We may venture into particle physics, lasers, anti-matter, space travel, biology, physics, botany, time travel, etc."

Part of the appeal of Sci-Fi Movie Nights is the selection of movies, which include some of the campiest sci-fi films ever made – the ones that are so bad they're actually good. Among the films that have been shown are the 1957 cult classic Plan 9 from Outer Space and Robot Monster, both widely considered contenders for the worst movies ever made. Plan 9 from Outer Space, for example, is advertised as "almost starring Bela Lugosi" because the famed actor died just four days after filming began, and so a stand-in was hired who spent the rest of the film with a cape over his face so that viewers wouldn't know that it wasn't really Lugosi. In the low-budget classic Robot Monster, Earth is visited by aliens played by actors wearing a bizarre combination of gorilla suits and diving helmets. The cheesiness factor is enormous, but that's all part of the fun, and it keeps the audience coming back for more.

Beyond sheer entertainment value, however, these films are also a fun and effective way of stimulating public curiosity about astronomy. In Plan 9 from Outer Space, for example, Earth is invaded by surprisingly human-looking aliens who have come to our planet out of concern that we humans are developing a weapon that could destroy the universe. This provides an excellent starting point for discussing a wealth of topics after the movie, such as interstellar distances, the challenges of space travel, and the likelihood that aliens from another world would look human.

Public response to the Sci-Fi Movie Nights has been incredible, with standing-room-only crowds of all ages. The program has also been successfully taken on the road to Arizona, Colorado, Washington, DC and California, and would likely be a hit most anywhere.

Let me give another example of the power of film for communicating astronomy in a different cultural context. In 1998, I taught the first university astronomy course ever in the Gambia, an impoverished nation in western Africa. The introductory course covered the usual topics such as our solar system, stars, galaxies, cosmology, and the search for life in the universe. My students were a delight to teach, and they asked many excellent questions. They also taught me much about western African cultures and Islam, and I learned as much or more from them as they did from me. Yet I have no doubt that the one thing they all remember from my course to this day is this: I showed them the movie E.T. The Extraterrestrial, which I had brought with me. Because of the extreme poverty in the Gambia – it is ranked as one of the poorest countries in the world by the United Nations – most of my students didn't have televisions and had never seen this movie before. And they loved it.

Toward the end of the film, E.T. and his young earthling friend, Elliot, are trying to escape from government agents when suddenly their bike flies into the air as if by magic, over a roadblock, over houses, and returns E.T. to his waiting spaceship. It's a wonderful moment in cinematic history – and my entire class burst into spontaneous applause. Steven Spielberg is, of course, a master at pushing those universal emotional buttons that people of all cultures can relate to. After showing E.T. The Extraterrestrial in my class, I had a captivated group of students for our ensuing discussion of the search for life in the universe. The possibility of extraterrestrial life had suddenly been transformed from an abstract concept into an alien creature with a face, feelings, and a name – even if it was just fantasy and highly anthropomorphized, a personal connection to a scientific question had been created.

Finally, let me also mention the possibility of piggybacking off of major Hollywood releases to increase public understanding of astronomy. When the new sci-fi comedy movie The Hitchhiker's Guide to the Galaxy was released, in which the Earth is destroyed to make way for an intergalactic hyperspace bypass, the University of Hawaii's Institute for Astronomy took advantage of the buzz generated by the film by offering its own free evening presentation titled "The Hitchhiker's Guide to the End of Everything" in which a panel of astronomers discussed the chances of surviving real astronomical dangers, such as killer asteroids, colliding galaxies, exploding stars, black holes, and the eventual fate of the Earth and the universe. It was a timely and fun way of communicating science. New movies with some astronomical connection are released all the time (e.g., the recently released War of the Worlds), providing new opportunities for capitalizing on the free publicity to increase awareness of astronomy.

**Conclusions**

My goal here has been to showcase some of the many non-traditional ways that the astronomical community is reaching out to communicate with the public. Many of these are quite inexpensive, and most can be quite fun. It is my hope that some of the ideas presented here will inspire others to find their own new and creative ways of communicating astronomy to people of all ages.

The original webcast of this talk, and all the talks at the Communicating Astronomy with the Public 2005 conference, can be viewed on the conference web page at www.communicatingastronomy.org/cap2005/

**Acknowledgments**

I thank the conference organizers for a most enjoyable meeting, and the European Southern Observatory for its hospitality and kind support.